\documentclass[jcp, preprint,
]{revtex4-2}

\usepackage{graphicx}
\usepackage{dcolumn}
\usepackage{bm}
\usepackage{amsmath}
\usepackage{amsfonts}
\usepackage{amssymb}
\usepackage{bm}
\usepackage{xcolor}
\usepackage{braket}
\usepackage[normalem]{ulem}
\usepackage{bold-extra}

\usepackage{todonotes}

\newcommand{\codename}[1]{\texttt{#1}}
\newcommand{\langname}[1]{\textsc{#1}}

\begin{document}

\title{Integrated workflows and interfaces for data-driven semi-empirical electronic structure calculations}

\author{Pavel Stishenko}
\email{These authors contributed equally to this work.}
\affiliation{Cardiff Catalysis Institute, School of Chemistry, Cardiff University, Park Place, Cardiff CF10 3AT, United Kingdom}
\author{Adam McSloy}
\email{These authors contributed equally to this work.}
\affiliation{Warwick Centre for Predictive Modelling, School of Engineering, University of Warwick, Coventry, CV4 7AL, United Kingdom}
\author{Berk Onat}
\affiliation{~Warwick Centre for Predictive Modelling, School of Engineering, University of Warwick, Coventry, CV4 7AL, United Kingdom}
\author{Ben Hourahine}
\email{benjamin.hourahine@strath.ac.uk}
\affiliation{SUPA, Department of Physics, John Anderson Building, 107 Rottenrow, University of Strathclyde, Glasgow G4 0NG, United Kingdom}
\author{Reinhard J. Maurer}
\affiliation{Department of Chemistry, University of Warwick, Coventry, CV4 7AL, United Kingdom}
\affiliation{Department of Physics, University of Warwick, Coventry, CV4 7AL, United Kingdom}
\author{James R. Kermode}
\email{j.r.kermode@warwick.ac.uk}
\affiliation{~~Warwick Centre for Predictive Modelling, School of Engineering, University of Warwick, Coventry, CV4 7AL, United Kingdom}
\author{Andrew Logsdail}
\email{LogsdailA@cardiff.ac.uk}
\affiliation{~Cardiff Catalysis Institute, School of Chemistry, Cardiff University, Park Place, Cardiff CF10 3AT, United Kingdom}

\date{\today}%

\begin{abstract}
Modern software engineering of electronic structure codes has seen a paradigm shift from monolithic workflows towards object-based modularity. Software objectivity allows for greater flexibility in the application of electronic structure calculations, with particular benefits when integrated with approaches for data-driven analysis. Here, we discuss different approaches to create deep modular interfaces that connect big-data workflows and electronic structure codes, and explore the diversity of use cases that they can enable. We present two such interface approaches for the semi-empirical electronic structure package, \codename{DFTB+}. In one case, \codename{DFTB+} is applied as a library and \textit{provides} data to an external workflow; and in another, \codename{DFTB+} \textit{receives} data via external bindings and processes the information subsequently within an internal workflow. We provide a general framework to enable data exchange workflows for embedding new machine-learning-based Hamiltonians within \codename{DFTB+}, or to enabling deep integration of \codename{DFTB+} in multiscale embedding workflows. These modular interfaces demonstrate opportunities in emergent software and workflows to accelerate scientific discovery by harnessing existing software capabilities.\end{abstract}
\maketitle

\newpage
\section*{\label{sec:significance}Statement of Significance}
Software is integral to modern science, and modularity and interoperability of software will allow new capabilities that are greater than the sum of the constituent elements. We demonstrate such capabilities via the integration of modular software packages with the semi-empirical electronic structure package \codename{DFTB+}. Our work is exemplar in showing how a well-developed software package, such as \codename{DFTB+}, can undergo re-engineering to allow work package and workflow interoperability in either client or server capacity. It is imperative that the community of chemical physics embraces the opportunities presented by modular software design and interoperability of software packages; our work is therefore significant because it provides original demonstrations of how modular software integration can lead to higher-value outcomes from electronic structure calculations. These new capabilities can be used to guide further developments in this field and related applications.

\newpage
\section{\label{sec:intro}Introduction}
Semi-empirical electronic structure methods, such as Density Functional Tight-Binding (DFTB) theory,\cite{doi:10.1098/rsta.2012.0483, bannwarth24:_semiem} have a long-standing history of enabling fast and robust predictions on a diverse range of materials for time and length scales.\cite{bannwarth24:_semiem} The atomistic resolution accessible with DFTB (up to $\sim10^{18}$ 
atoms)\cite{nishimura2021quantum} is traditionally out of reach for conventional first-principles calculations, making these approaches particularly appealing for large-scale simulation of dynamical chemical processes. Semi-empirical approaches can provide robust accuracy for conventional organic molecular materials\cite{Gaus2013, mortazavi_structure_2018} or inorganic materials.\cite{jha_dftb_2022}
Integration of these approaches in complex automated computational workflows and machine learning (ML) surrogate models is now timely given the impact of these new capabilities on computational materials science and chemistry over the last decade.\cite{keith_combining_2021,westermayr_perspective_2021,fedik_synergy_2023} 

The concept of using data-driven approaches to transfer first-principles information into second-principles electronic structure codes has a long history in the construction of tight binding parametrizations; for example, through global stochastic optimization of tight-binding parameters or repulsive potentials via particle swarm optimization.\cite{knaup2007initial,aguirre_development_2020, hutama_density-functional_2021} With the emergence of modern ML methods, the prospect of closer integration and direct learning between methods has been explored by several studies. As an example, St\"ohr \textit{et al.} have used ML interatomic potentials to represent the repulsive potential in DFTB;\cite{stohr_accurate_2020} and several studies have shown that ML surrogate models can accurately predict first-principles electronic structure in local orbital representation.\cite{Schutt2019, gastegger_deep_2020, unke_se3-equivariant_2021, nigam_equivariant_2022, zhang_equivariant_2022, li_deep-learning_2022} In the context of Density Functional Theory (DFT) and other semi-empirical methods, ML has also been used to represent electronic Hamiltonian parameters, \cite{Dral2015, zhou_deep_2022} including the \codename{TBmalt} approach providing end-to-end learning of parameters based on target properties.\cite{mcsloy_tbmalt_2023} The proof-of-principle applications show what is possible in this space, but standardized workflows or integrated approaches are yet to emerge.

The majority of well-established electronic structure software packages have developed as monolithic codebases with a single entry point, due to extensive investment in their development before the widespread adoption of integrated workflows. The consequence is a bottleneck for the integration between electronic structure and big-data approaches. Electronic structure software packages with modularity in their design, that provide external accessibility to inner functionality, have become more prominent in recent years, reflecting the evolving landscape of computational materials science and the uptake of objective, modular programming and data-driven workflows (\textit{e.g.}, \codename{GPAW},\cite{Mortensen2005-dv} \codename{Psi4},\cite{smith2020psi4} and \codename{pySCF}\cite{sun2020recent}) or UNIX philosophy\cite{mcilroy1978unix} inspired designs such as \codename{WIEN2k}\cite{10.1063/1.5143061}; furthermore, established packages have sought to reshape their designs with library components that allow execution through an externally-driven interface. These retrofitted approaches typically rely on file input/output (I/O) and parsing, with only basic variable communication (\textit{e.g.}, MPI communicators), though alternative strategies with in-memory data transmission have increased in popularity. 

Examples of established strategies include the automated building of deep \langname{Python} interfaces to Fortran codes using \codename{f90wrap}\cite{kermode2020f90wrap} \textcolor{black}{(\textit{e.g.}, \codename{QEPy} for \codename{Quantum Espresso} \cite{qepy}, \codename{quippy} for  QUIP \cite{Csanyi2007QUIP}, CasPyTep for \codename{CASTEP}\cite{Clark2005-dx}, and a Python wrapper for the Bader code \cite{f90wrap_Bader})}, socket-type interfaces \textcolor{black}{(i-PI \cite{KAPIL2019214}, MDI \cite{MDI2022})}, and library extraction \textcolor{black}{(\codename{tblite} in \codename{CREST} \cite{CREST2024}, ELSI \cite{YU2020107459})}. High-level Python packages, like the Atomic Simulation Environment (ASE)\cite{larsen2017atomic}, have emerged as a \textit{de facto} standard for building \textcolor{black}{frameworks for} atomistic simulation workflows. \textcolor{black}{Examples of workflow engines include \codename{AiiDA} \cite{pizzi2016aiida}, \codename{Chemoton} \cite{CHEMOTON2}, \codename{CENSO} \cite{CENSO}, \codename{BigChem} \cite{BigChem}, \codename{QCEngine} \cite{MDI2022}, \textit{etc.}, which enable} some classes of high-level algorithms to be written once and reused between codes; however, this generality is commonly restricted to atomic and molecular properties rather than electronic \textcolor{black}{structure}. 

Recently, the emergence of automated and machine-learning-augmented workflows, and establishment of extensive materials databases using FAIR principles,\cite{wilkinson2016fair} has led to a need for more flexible infrastructure capabilities in the design of electronic structure software. In particular, there is evidence of a clear need for greater modularity and interoperability in code design, which should support strong interfaces between electronic structure codes and external software packages. Such developments would circumvent traditional bottlenecks in data communication and accelerate discoveries facilitated by electronic structure theory.

Currently, there remains limited demonstration and standardization of deployable, user-ready interfaces that facilitate interaction and application of external workflows and data-driven frameworks with first-principles and semi-empirical electronic structure software packages. In particular, there is a need for deep module interfaces that can expose the extensive and well-developed functionality in existing established software. The interfaces should be simple code-level interfaces, rather than commonly shallow interfaces that work predominantly with complex file I/O or external scripted workflows. The ability to use such deep interfaces, which allows large data objects to be manipulated during run-time, will reward the community with computationally efficient approaches concerning both data processing and data storage. \textcolor{black}{The list of software that may benefit from using deep interfaces includes charge partitioning codes \cite{Bader, DDEC6_part1, Chargemol}, electronic structure analysis \cite{wannier90}, and the growing family of electronic structure machine learning models \cite{zhang_equivariant_2022, mcsloy_tbmalt_2023, Dral2015, li_deep-learning_2022}.} The Atomic Simulation Interface (ASI)\cite{JOSS} is a recent example interface that is built to import and export electronic structure information from quantum chemistry codes during runtime with minimal performance penalty, as demonstrated \textit{via} coupling with the \codename{DFTB+} and \codename{FHI-aims} software packages.\cite{blum2009_fhiaims}

The commonly articulated strategies for integration of electronic structure software packages can be categorized in order of depth and complexity of the interface as:
\begin{description}
\item[Data parsing via file I/O operations] typically focused on input and output data files only. This provides no data accessibility at the mid-process stage and has limited data precision and data object sizes.
\item[Socket (and alternative) data transport protocols] small data objects are communicated in byte format \textit{via} a lower-level transport method.\cite{KAPIL2019214, MDI2022} This provides data-accessibility mid-process but is limited in terms of data object size.
\item[Directly connecting to an API] an Application Programming Interface (API) provides a well-defined set of function calls. This provides data accessibility mid-process, can handle flexible data object sizes, and provides fixed API standards.
\item[Flexible interfacial wrappers] an intermediate package manages couplings between different API standards across multiple languages. This provides data-accessibility mid-process and is flexible in both data object size and API definitions.
\end{description}

\textcolor{black}{Despite the recognised disadvantages of file I/O-based interfacing, this approach is the most commonly used option. Reasons that file I/O interfacing remains popular is that the file system interface is universally provided by the operating system, it does not impose any restrictions on data format, and it is intrinsically asynchronous. These features are useful if codes on both ends of the interface are not specified \textit{a priori}, which is especially applicable in the case of closed-source software. Motivations for deeper interfaces include the complexity of I/O data formats and the stability of their layout over time, the synchronisation of I/O data, and inherent performance limitations when using I/O, especially at high frequency.} \textcolor{black}{The ongoing efforts towards standardisation of input and output formats, such as \codename{QCEngine} \cite{MDI2022}, JSON \cite{pezoa2016foundations} in ORCA \cite{ORCA} or HDF5-based TREXIO \cite{TREXIO}, offer potential to alleviate some of these challenges.}

Recent work on the \codename{DFTB+} software package\cite{dftb+, 10.1063/5.0103026} has investigated how deep interfaces can be established and used for the benefit of workflow-based computational simulation. Herein, we discuss general strategies for interfacing to electronic structure codes before presenting two interfaces that are capable of either being \textit{driven by an external workflow} to provide data,\cite{JOSS} or \textit{driving a workflow}, with external bindings used to obtain data from an external engine.\cite{zhang_equivariant_2022} The interfaces follow different strategies and philosophies, and address distinct potential use cases, such as embedding and ML workflows.

\section{\label{sec:interfaces}General considerations on deep interfaces to electronic structure codes}

\begin{figure*}[!hbt]
\includegraphics[width=0.8\textwidth]{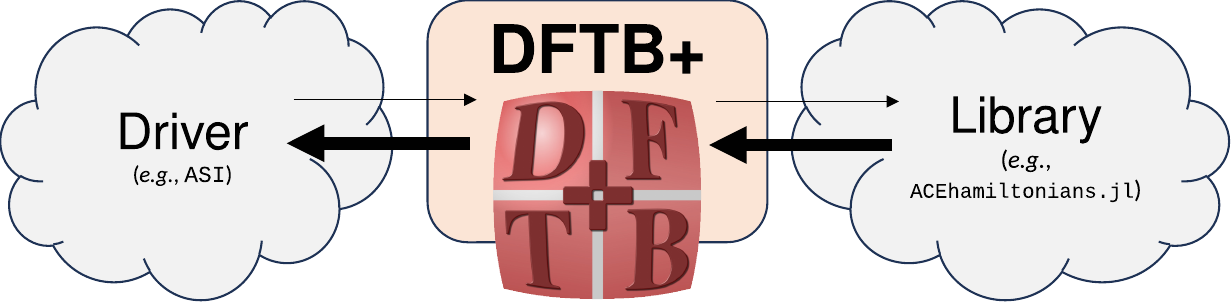}
\caption{Schematic representation of interfacing paradigms where \codename{DFTB+} is used: as a resource, driven by an external package (left-hand side); or instead drives communication with an external package as part of its own workflow (right-hand side). In both cases, the majority of data communication is returned to the software driving the relationship (\textit{i.e.}, client), as indicated by the asymmetry in the arrows representing data flow.}
\label{fig:paradigm}
\end{figure*}

The contemporary difficulties of interfacing with electronic structure software packages are rooted in the assumption that the transient data objects that describe the electronic structure of a simulated system are not valuable outside of the code. Therefore, such data structures were often placed deep in the code foundations, and exposing these data structures for read or write operations typically requires intrusive changes in the core codebase.

An additional difficulty when interfacing with electronic structure software packages is the substantial size of electronic structure descriptors (electronic density, Kohn-Sham orbitals, matrices of Kohn-Sham equations). Simulations often take up the majority of available random access memory and transferring large amounts of data between formats, or copying between codes, can be limited by accessible memory and become a performance bottleneck.
Unnecessary copying of the data between inter-operating codes can be avoided by providing direct access to the memory buffers \textit{via} shared memory, memory-mapped files, or by running these codes as libraries in a single process; however, such approaches face other obstacles caused by data efficiency practices in high-performance computing, such as the reuse of large arrays for storage of various different data. For example, many routines in the Basic Linear Algebra Subprograms (BLAS) library return their results in their input buffers,\cite{blackford2002updated} overwriting the data in repeated execution; if one wants to, \textit{e.g.}, export a large array, a pointer to the internal data buffer is insufficient and instead the data must be accessed when the desired data is actually in that array. In practice, accessing the data in this manner means that the code execution must be paused and restarted at various (initially unplanned) moments, often deep in the call stack. Given that many electronic structure software packages have been initially designed as standalone applications, such changes in the control flow can be intrusive and error-prone.

Approaches to suspend and restart a subroutine map on to two different code interfacing paradigms (see Figure \ref{fig:paradigm}). One option is refactoring and splitting code into two separate subroutines that are called immediately before and after any data import or export. An alternative option is invoking a user-provided callback function to perform data read or write. The former method essentially converts the code into a library, inverting the control of the workflow; if the routine that is split is nested deep within the call stack, every function along the call stack must also be split. The latter method introduces local inversions of control through callbacks, but the modified code generally still drives the overall workflow. 

\textcolor{black}{In general, the considerations outlined may be viewed as significant drawbacks to deep interfaces, especially if the interfaced code is poorly structured, fragile, or tightly coupled; in such cases, development and maintenance of deep interfaces may become involved and laborious. However, the potential benefits are significant and, if pursued carefully, present opportunities for realising new science.} In the following, we present realisations of \textcolor{black}{the two highlighted} deep interfacing paradigms inside the \codename{DFTB+} software package\textcolor{black}{, outlining our experiences with realisation and showcasing the potential value.}

\section{Methods}

\subsection{The DFTB family of models}
The DFTB method,\cite{doi:10.1098/rsta.2012.0483} and related
semi-empirical tight-binding models,\cite{bannwarth2020} approximate density functional theory (DFT). By expanding the Kohn-Sham functional around an approximate reference density, the
total energy expressions are written as a sum of: a (generally) attractive electronic band-structure contribution; an electrostatic energy; and a repulsive energy (which corresponds to the double-counting terms in DFT). The expressions for the electrostatic contributions are derived with respect to fluctuations from a reference charge density, which itself is assumed to be the sum of a set of neutral atomic densities corresponding to the structure being
modeled. The electronic structure Hamiltonian itself is typically evaluated from reference neutral atoms and atomic dimers. The 2-centre integrals are for a minimal, non-orthogonal, atomic valence basis (neglecting crystal field and 4-centre contributions\cite{doi:10.1098/rsta.2012.0483}).  Depending on the choice of Hamiltonian (DFTB or xTB) these values are (typically) obtained from DFT calculations, either by tabulation or by fitting empirical expressions.

The charge fluctuations from the neutral reference are expressed using Mulliken (gross) charges\cite{10.1063/1.1740588} and, depending on the
Hamiltonian, the electrostatics are restricted to atomic monopoles or selected multipole contributions. The electrostatic
potential is then evaluated at each atomic site, with the resulting 2-centre contributions approximating the integrals as a product of the overlap between sites and the average of their potentials (for the monopole).

The exchange-correlation contributions are included in the
parameterization of the reference neutral system, combined with taking a suitable atomic limit for the electrostatic energy of the charge fluctuations.\cite{Elstner1998,doi:10.1021/acs.jctc.8b01176} The double-counting terms in the energy expressions are represented as fitted inter-atomic potentials\cite{doi:10.1098/rsta.2012.0483,dftb+} or as parameterized inter-atomic integrals.\cite{bannwarth2020}

\subsection{\label{sec:dftbp} The \codename{DFTB+} code}

The \codename{DFTB+} code implements various DFTB and xTB models. Interactions between atoms are internally represented using a data structure based on spatial atomic neighbors\cite{aradi2007dftb+} and most terms are evaluated in real space, hence the majority of the code is boundary-condition independent. Therefore, for periodic structures (and other space-filling geometries), the Hamiltonian and overlap matrices are transformed into a crystal-momentum ($\mathbf{k}$) dependent dense Hamiltonian. The resulting set of secular equations for the band structure is solved either \textit{via} conventional diagonalization (LAPACK\cite{laug} or ScaLAPACK\cite{slug}), \textit{via} hybrid CPU-GPU calculations (MAGMA\cite{dghklty14, vuong2023accelerating} or ELPA\cite{YU2021107808}) or through one of the eigenvalue or density-matrix distributed solvers provided by the \codename{ELSI} project.\cite{YU2020107459}

The \codename{DFTB+} codebase\cite{hourahine_2023_8117766} is
primarily written in \langname{Fortran 2008}, with components in
\langname{C}/\langname{C++} and \langname{Python3}. An API is provided in these languages to use the code as an external library for energy/force calculations, or other modes such as real-time electronic propagation, \textit{e.g.}, Ehrenfest dynamics,\cite{bonafe2020real}). The software is licensed under the GNU Lesser General Public License 3.0 (or
later),\cite{LGPL3} chosen based on the code's library capability.
Continuous integration of \codename{DFTB+} is performed via the
project GitHub repository,\cite{hourahine_2023_8117766} with custom regression testing scripts, plus a unit test system using the \codename{FyTest} framework.\cite{fytest} The code is internally documented with \codename{Doxygen}\cite{doxygen} and
\codename{FORD}\cite{ford} compatible comments.

\subsection{Extensions enabled by the present work}

The external Hamiltonian evaluation via ACEhamiltonians.jl\cite{zhang_equivariant_2022} (Section~\ref{sec:JL}) is performed in real space, hence can be evaluated for the general range of boundary conditions supported by \codename{DFTB+}. These include conventional molecular/cluster structures in free space or periodic boundary conditions; more general boundary conditions can also be evaluated by \codename{DFTB+}, such as Green's function embedding\cite{Pecchia_RPP} or helical structures,\cite{10.1063/1.4819910}. The externally provided electronic structure model is built piece-wise from local geometric cluster fragments to give coverage of the entire geometry, that then includes the boundary conditions managed by \codename{DFTB+}.

The \codename{ASI} bindings (Section~\ref{sec:ASI}) directly exchange the dense Hamiltonian matrices to be diagonalized, and/or the dense single-particle density matrix, between codes. The direct communication enables direct comparison of the semi-empirical Hamiltonians against local or non-local first principles models. The local potential exchange via \codename{ASI} (Section~\ref{sec:ASI-electrostat}) also enables the use of various forms of external electrostatic embedding models, along with testing of the approximations in self-consistent semi-empirical Hamiltonians against the first principles local potentials.

\subsection{Computational Details}

To demonstrate the capabilities of the \codename{ASI} interface, we have calculated the band structure of Al with the Hamiltonian ($\mathbf{H}$) and overlap ($\mathbf{S}$) matrices evaluated in the all-electron full-potential numerical atomic orbital software package \codename{FHI-aims} (Version: 230905) \cite{blum2009_fhiaims} that implements ASI API version 1.1.\cite{JOSS} The ground state electron density of the Al bulk crystal with a lattice parameter of 2.024 \AA was evaluated using DFT with the PBE exchange-correlation functional,\cite{perdew1996generalized} a scalar-relativistic zeroth order regular approximation (ZORA) correction,\cite{blum2009_fhiaims} and a $2\times2\times2$ $\Gamma$-centered $\mathbf{k}$-grid. A ``minimal'' basis set was used for Al, which consists of 13 numerical orbitals (3s, 2p, and 1d orbitals). $\mathbf{H}$ was subsequently evaluated along the $\mathbf{k}$-path $W$-$X$-$\Gamma$-$L$-$K$-$\Gamma$-$L$, with 50 sampling points in each section of the pathway. $\mathbf{H}$ and $\mathbf{S}$ were exported via the \codename{ASI} callback functions from \codename{FHI-aims} to \codename{DFTB+}. The \codename{DFTB+} computation was configured to use the same basis and path in $\mathbf{k}$-space, with the eigensolver of \codename{DFTB+} used to obtain the band structure.

In the case of the \codename{ACEhamiltonians interface}, the same \codename{FHI-aims} configurations were used. The $\mathbf{H}$ and $\mathbf{S}$ matrices were exported from \codename{FHI-aims} as real-space matrices using \codename{ACEhamiltonians} version v0.1.0.\cite{zhang_equivariant_2022}. 

\section{\label{sec:JL}\codename{ACEhamiltonians} as a library to provide external Hamiltonian evaluation for \codename{DFTB+}}

Within \codename{DFTB+}, an interface was constructed to facilitate communication between the \codename{ACEhamiltonians} and \codename{DFTB+} software packages. The interface enables data-driven \codename{ACEhamiltonans} models for \textbf{H} and \textbf{S} to be combined with the robust functionality of the \codename{DFTB+} framework. The interface design provides threefold benefit: (i) modularity, by providing a means by which observables can be computed using \codename{ACEhamiltonians} models without having to unnecessarily extend the \codename{ACEhamiltonians} codebase itself; (ii) performance, by using an optimized production-level framework such as \codename{DFTB+}, especially when dealing with larger systems where domain decomposition-based parallelism is essential; (iii) accessibility, as such an interface reduces the barrier associated with using an \codename{ACEhamiltonians} model by allowing combination with widely used software that has a broad userbase.

\subsection{\label{sec:JL-methods} Interface Description}

\begin{figure*}[!htp]
\includegraphics[width=0.8\textwidth]{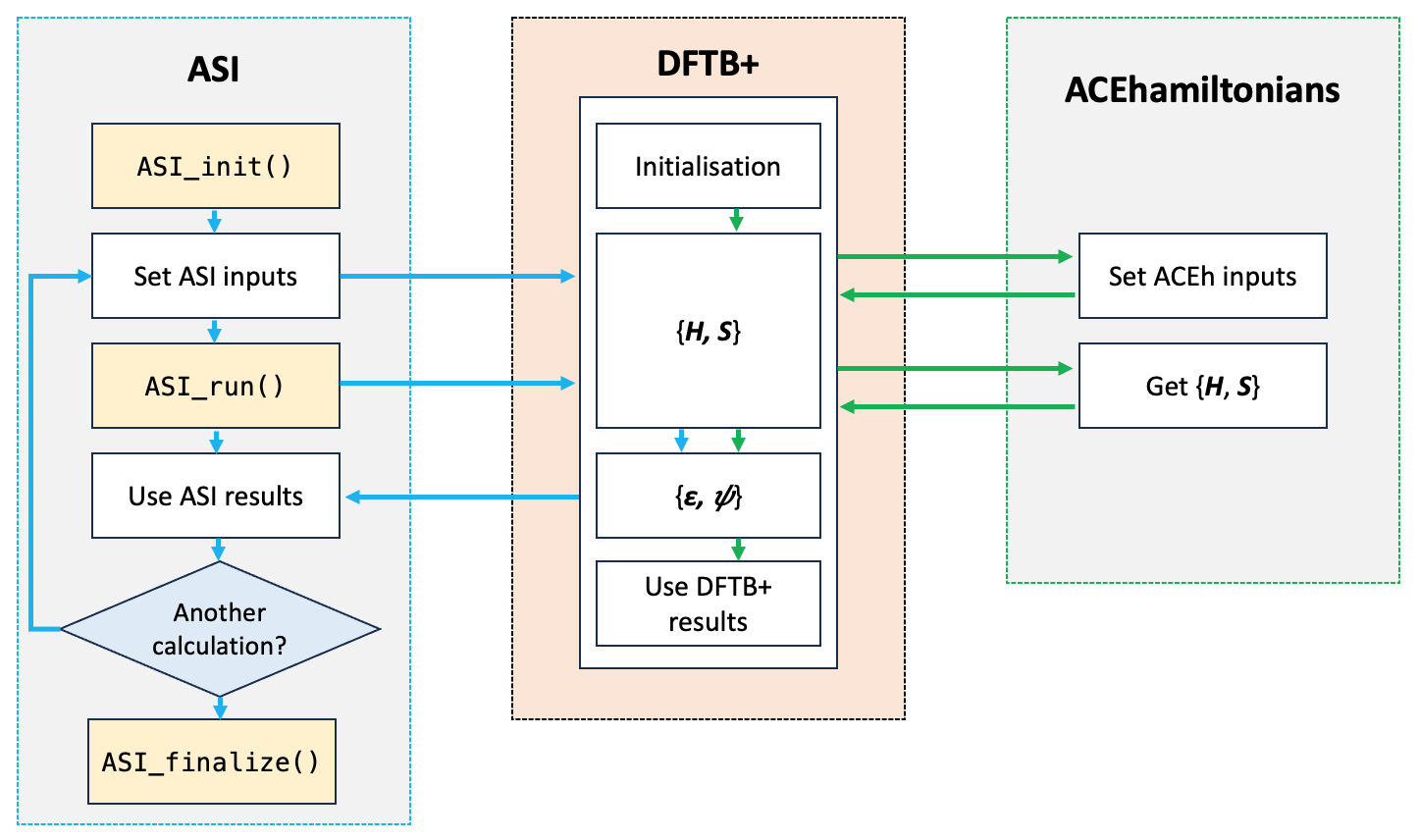}
\caption{Schematic representation of the specific workflows invoked between \codename{DFTB+}, \codename{ASI}, and \codename{ACEhamiltonians}. Boxes and arrows in blue are part of the \codename{ASI} execution pathway, and in green for the \codename{ACEhamiltonians} pathway}
\label{fig:schematic_detailed}
\end{figure*}

Communication between the \langname{Fortran}-based \codename{DFTB+} and \codename{ACEhamiltonians} 
 follows the general structure illustrated in Figure \ref{fig:schematic_detailed}, except that the \langname{Julia}-based \codename{ACEhamiltonians} is facilitated via an intermediary \langname{C} layer. This interfacial layer ensures that the modifications to \codename{DFTB+}, which allow invocation of external models, are not restricted to one external framework or programming language. The translation layer was written in low-level \langname{C}, which provides good interoperability with other languages through external bindings. When executed, \codename{DFTB+} calls the \codename{ACEhamiltonians} interface to invoke a setup subroutine, during which an initial bidirectional exchange of information occurs. The exchange allows \codename{DFTB+} to specify the chemical species present in the target system; the \codename{ACEhamiltonians} interface responds by providing the environmental and interaction cutoff distances, followed by the number of orbitals present for each species along with their occupancy and azimuthal quantum numbers.

\codename{DFTB+} constructs all relevant atom and bond clusters with the information obtained. The clusters are provided to the interface, along with a list of indices specifying which block of the Hamiltonian/overlap matrix are represented by each cluster. The information is stored internally in the \codename{ACEhamiltonians} interface until a new set of clusters is provided, such as would be expected during a molecular dynamics simulation; or the model cleanup subroutine is invoked, which clears memory in preparation for code termination.

Subsequently, \codename{DFTB+} calls the prediction subroutine in the \codename{ACEhamiltonians} interface, providing pointers to the Hamiltonian and overlap matrices that are to be populated. The interface loops over the atom clusters and populates the associated on-site Hamiltonian matrix block by block for each atom by evaluating the model. During each loop, the \codename{ACEhamiltonians} function responsible for constructing on-site blocks is called; the coordinates and species of the atoms are provided, along with the model that is to be evaluated and the block of the Hamiltonian matrix into which the results should be placed; the on-site blocks of the overlap matrix are set to an identity matrix. The process is repeated with the bond clusters to fill in the offsite blocks of the Hamiltonian and overlap matrices.

\begin{figure}[!ht]
\includegraphics[width=0.8\textwidth]{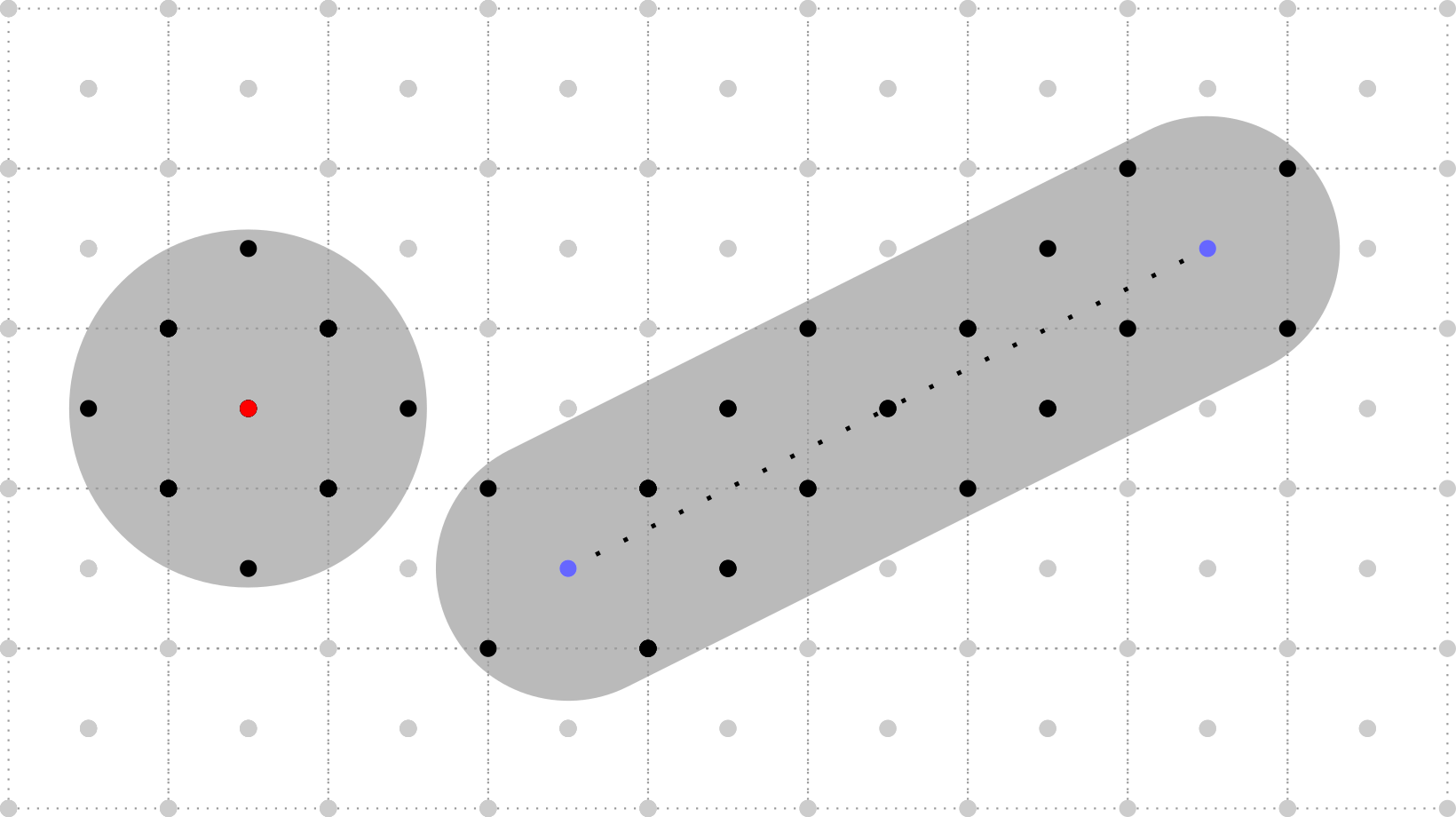}
\caption{Schematic representation of an atom-centred cluster (central atom shown in red), and a bond-centred cluster formed between a pair of atoms (shown in blue), in a periodic crystal lattice. Shaded regions indicate areas where an atom would be considered to be part of the cluster. Black and gray colored atoms are used to indicate those that are and are not part of a cluster, respectively.}
\label{fig:clusters}
\end{figure}

As shown in Figure \ref{fig:clusters}, the clusters needed to compute onsite blocks are spherical and atom-centered, while those for offsite blocks, which represent interactions between orbitals on distinct two atoms, are cylindrical and bond-centered. Atomic coordinates are provided relative to the origin atom $i$, with which atomic clusters are constructed for every atom in the structure. The atomic cluster for atom $i$ can be defined as the subset of atoms that satisfy $r_{ij} \le r_{cut}$; where $r_{ij}$ is the distance between atom $i$ and some other atom $j$, and the environmental cutoff distance $r_{cut}$ is a free parameter. Bond clusters are created for all atom pairs $\{i,j\}$, for which atom $i$ resides in the origin cell and $r_{ij} \le r_{bond}$ holds true; where $r_{bond}$ specifies the interaction cutoff distance. For a given atom pair $\{i,j\}$, the bond cluster is the subset of atoms whose perpendicular distance to the open line segment between atoms $i$ and $j$ does not exceed the specified environmental cutoff distance $r_{cut}$. All coordinates are specified to the midpoint of the bond. Further details of the \codename{DFTB+} external API are given in the Supporting Information (SI).

\subsection{Results}
\begin{figure}[!ht]

\includegraphics[width=0.8\textwidth]{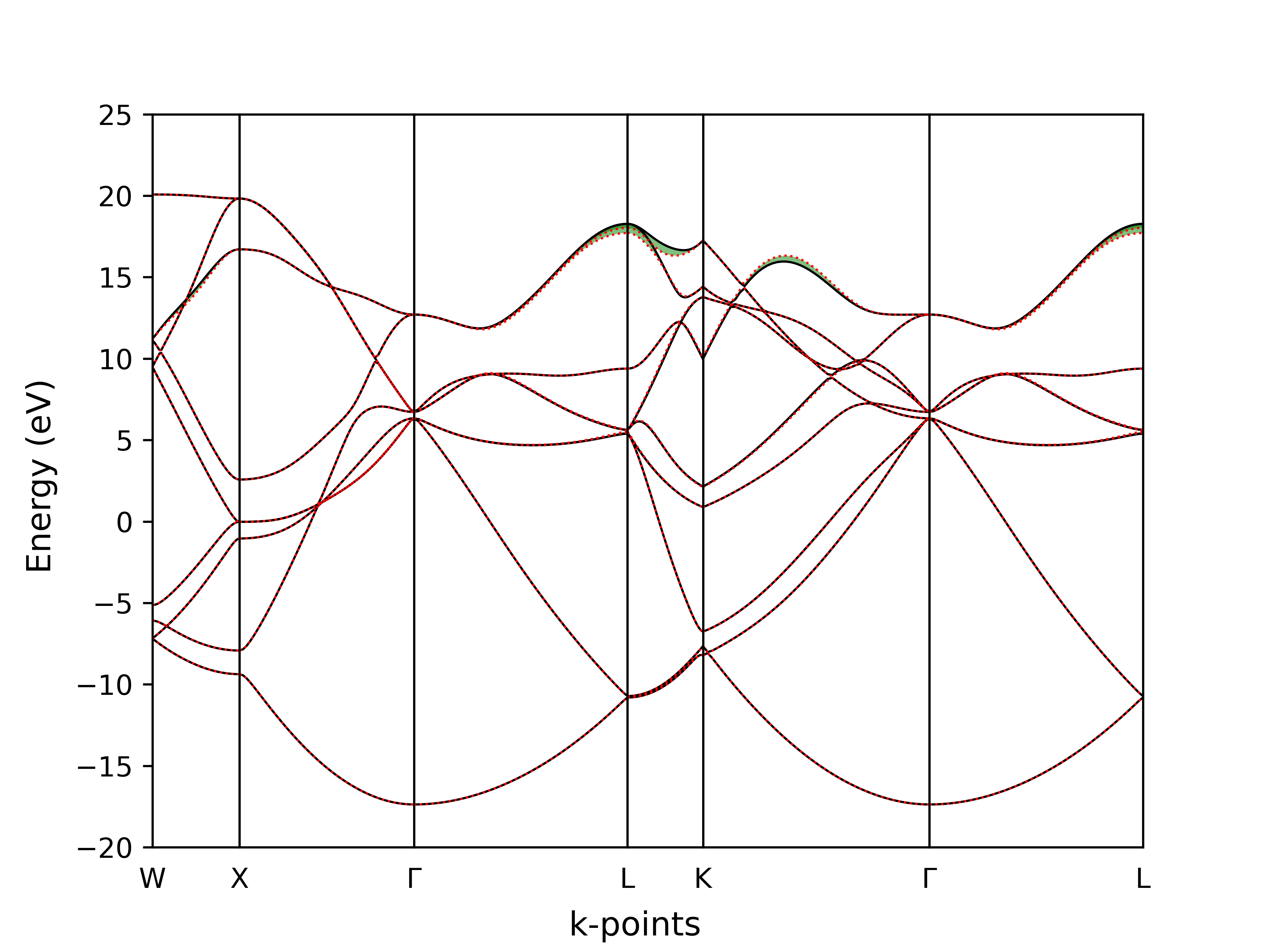}
\caption{Band structure of a pristine aluminum FCC unit cell as calculated directly by the \codename{ACEhamiltonians} package (black), and obtained using the \codename{DFTB+} API with the same model (red dotted). Green shading is used to highlight areas of discrepancy in the unoccupied states.}
\label{fig:fcc_band_acehamiltonians}
\end{figure}

Figure~\ref{fig:fcc_band_acehamiltonians} presents the band structure of a pristine aluminum FCC unit cell as obtained \textit{via} the \codename{DFTB+} API (red dots), alongside the same calculation performed using the \codename{ACEhamiltonians} package directly (black line). The results agree quantitatively in the occupied levels, however local deviations become apparent within the higher energy unoccupied levels. Notably, the band structure exhibits a mean absolute eigenvalue error of approximately $10^{-2}$~eV. This is greater than would naturally be expected given that the two results are generated using the same underlying model, and share many of the same prediction subroutines. In an effort to determine the source of the observed deviation, the \codename{DFTB+} API was used to generate and subsequently write out the Hamiltonian and overlap matrices. These were then used by \codename{ACEhamiltonians} to reconstruct the band structure.  The resulting band structure was within machine precision of that derived using \codename{ACEhamiltonians} directly, which demonstrates that the discrepancy originates from the means by which the band structures were calculated rather than the underlying matrices (\textit{i.e.}, the API is not the source of the difference). The discrepancy stems from the different eigensolvers used by the \codename{ACEhamiltonians} and \codename{DFTB+} codebase: when using a common matrix source, the subroutines of the \codename{ACEhamiltonians} package produce band structures that agree ($\sim8\times10^{-6}$~eV) with those generated by \codename{FHI-aims}, and alleviates the discrepancies in the higher energy levels.

\section{\label{sec:ASI} Atomic Simulation Interface (\codename{ASI}) as a driver that uses \codename{DFTB+}}

The ability to drive calculations externally, and use a specific package to evaluate system properties on demand, motivates the development of an infrastructure where \codename{DFTB+} can be deployed as a software library. Modern \langname{Python} coding developments provide capacity for high-level interfaces, reliant on file I/O for data transfer, but deep integration \textit{via} pre-compiled software languages can enable more efficient and accurate data communication and software application. Recent efforts towards this software paradigm have seen the development of the Atomic Simulation Interface (\codename{ASI}), with the primary purpose of conveniently connecting \codename{ASI}-enabled codes in multiscale simulation workflows, such as hybrid quantum/molecular mechanics (QM/MM), multiscale quantum mechanical embedding (QM/QM), or integration with machine learning (ML) frameworks (QM/ML).

\subsection{\label{sec:ASI-methods} Interface Description}


ASI has been developed as a plain \langname{C} API, again demonstrating the use of a low-level language enabling compatibility across software infrastructure. The key feature of the \codename{ASI} is the provision of an efficient and portable method to transmit large data arrays, relevant to electronic structure models, between software packages. \codename{ASI} itself is fundamentally an API specification, similar to MPI or BLAS standards, that ensures compatibility; the complete \codename{ASI} API specification is available as a \langname{C} header file with comments in \codename{Doxygen}\cite{doxygen} format, along with \langname{HTML} pages generated by Doxygen\cite{doxygen} from the aforementioned \langname{C} header file. The \codename{ASI} API is designed to be implemented by software packages to provide programmatic access to their internal data structures. We refer to software that implements \codename{ASI} API as \textit{ASI-enabled codes}, and we refer to software that invoke \codename{ASI} API functions as \textit{ASI clients}.

In the current example, \codename{DFTB+} is an \codename{ASI}-enabled code with functionality provided for the communication of key electronic data structures, such as Hamiltonian ($\mathbf{H}$) and overlap ($\mathbf{S}$) matrices, as well as less complex data objects, \textit{e.g.}, variables and arrays, such as energies ($E$) and forces on atomic centers ($-\nabla E$). The \codename{DFTB+} \codename{ASI} is implemented as a separate \langname{C} library that links with the \codename{DFTB+} library and \codename{ASI} clients. A \langname{Python} wrapper for the \codename{ASI} API, \codename{asi4py}, provides compatibility with \langname{Python} workflows and is available for installation \textit{via} the \codename{pip} command line tool. The convenience of \codename{asi4py} complements the deep integration of the \codename{ASI} interface and provides a user-friendly way to create \codename{ASI} clients in Python.

The key \codename{ASI} functions can be broken into four groups: control flow; atomic information; electrostatic potential; and electronic structure matrices.\cite{JOSS} The necessary intrusions to implement in an existing codebase are minimal. For the application of \codename{DFTB+} using the \codename{ASI} standard, the \codename{DFTB+} package is compiled as a shared object library to allow dynamic linkage with the client. The workflow is driven by the \codename{ASI} client; thus, after \codename{ASI} initialisation, key data objects are communicated to/from \codename{DFTB+} and callback functions registered before the request for execution of a \codename{DFTB+} calculation. Callback functions give direct access to \textcolor{black}{internal} data objects within \codename{DFTB+} \textcolor{black}{process \textit{via} pointers, without unnecessary copying,} thus causing near-zero computational \textcolor{black}{and communication} overhead, and also adapting to the chosen parallelisation scheme.  The callback functions are invoked during the execution process, providing external access to data objects when calculated. Derived quantities, such as energy, forces, stress, atomic charges, are also available. Once all necessary operations on the exposed data objects have been completed, finalization is performed, which includes the release of allocated memory. The \codename{ASI} workflow is presented in Figure \ref{fig:schematic_detailed}, and contrasted against the \codename{ACEhamiltonians} interface. Further details of the key \codename{ASI} functions are provided in the SI.

\subsection{\label{sec:ASI-results} Results}

\subsubsection{\label{sec:ASI-electrostat} Electrostatic embedding} 
\begin{figure}[ht]
\includegraphics[width=0.8\textwidth]{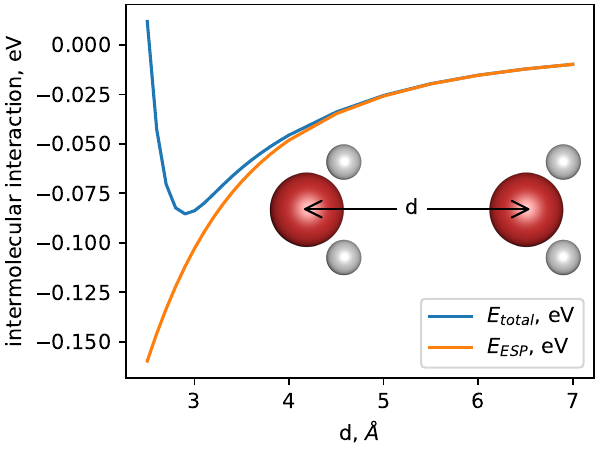}
\caption{Example of electrostatic embedding in DFTB+ achieved with the ASI interface. The distance (\textit{d}) between two water dimers, as shown in the inset. The graph shows the interaction energy as a function of \textit{d} (blue line), and the electrostatic embedding energy evaluated with \codename{ASI} API in a self-consistent manner (red line). The lines are shown to converge at large \textit{d}}
\label{fig:asi_h4o2}
\end{figure}

The \codename {ASI} functions that allow communication of the electrostatic potential can facilitate electrostatic QM/QM embedding. Figure \ref{fig:asi_h4o2} compares the total intermolecular interaction energy of two water molecules evaluated with \codename{DFTB+}, and separately the electrostatic component of that interaction as evaluated with a \langname{Python} script that orchestrates two \codename{DFTB+} instances using the \codename{asi4py} library interface. In the latter case, each \codename{DFTB+} instance calculates a single water molecule, and the electrostatic potential from one molecule is then exported from one \codename{DFTB+} instance, using \codename{ASI\_calc\_esp} function, and transferred \textit{via} MPI calls to the second \codename{DFTB+} instance, where it is included via the callback installed by the \codename{ASI\_register\_external\_potential} function. 

The calculation is performed self-consistently: the energy of both molecules is calculated with zero external potential initially; then the calculation for each molecule is repeated using the electrostatic potential provided by the other molecule. The calculation should be repeated until self-consistency is achieved. Convergence criteria should be defined and checked by ASI clients; for a simple system with two water molecules simulated by separate DFTB+ instances, five iterations are sufficient to reach $10^{-5}$ eV accuracy on the distances from 2.5 \AA  ~and above (see Figure \ref{fig:asi_h4o2}). Figure \ref{fig:asi_h4o2} shows that the electrostatic potential is dominant for the intermolecular interaction at large distance ($>4$ \AA), which is the expected behaviour.

\subsubsection{Electronic structure transfer} 
The \codename{ASI} implementation in \codename{DFTB+} supports the import of Hamiltonian ($\mathbf{H}$) and overlap ($\mathbf{S}$) matrices. With this functionality, data objects evaluated in electronic structure software packages can be imported into \codename{DFTB+} and evaluated. The potential of this functionality is demonstrated with computation of the electronic band structure for bulk Al (Supporting Information, SI, Figure S1), where $\mathbf{H}$ and $\mathbf{S}$ have been computed with the software package \codename{FHI-aims}. \codename{FHI-aims} supports the \codename{ASI} API, and with the \codename{ASI} data transfer protocols it is possible to calculate and analyse the band structure in \codename{DFTB+}. The resulting band structure is given in the SI: the result overlays the band structure achieved with the standalone \codename{ACEhamiltonians} approach and matches the \codename{FHI-aims} native calculation of the same data, showing the versatility of this modular interface.




\section{\label{sec:conclusions}Conclusions}

As workflows in computational materials science become more complex, codes need to become more interoperable. Potential paradigms when interfacing electronic structure software with other codes are: the software can act as the \textit{driver}, requesting information; or as the \textit{library}, being queried for information. In both cases, data transfer is bidirectional, although asymmetric. With the emergence of ML workflows, there are many opportunities to achieve synergy between semi-empirical electronic structure methods and data-driven approaches;\cite{fedik_synergy_2023} to yield usable software solutions in that enable complex simulations or data-driven workflows, robust interfaces between different codes must be established. 

Here, we have reported examples of electronic structure interfaces, implemented in the \codename{DFTB+} code, which explore the driver and library paradigms. We explain the general considerations and traits of the interfaces and showcase possible use cases by communicating electronic structure information in the form of the Hamiltonian in local basis representation, and evaluation of emebedding electostatic potential.

Both interfaces have the potential to provide exciting future capabilities. The \codename{ASI} bindings can in principle be used for a self-consistent workflow, either driven inside \codename{DFTB+} or externally. Similarly, the \codename{ACEhamiltonians} framework could exchange atomic properties such as charge, enabling self-consistent
updates of the supplied model. Either option would then also
immediately be compatible with a subset of the \codename{DFTB+} capabilities beyond ground state calculations, such as $\Delta$-SCF excitations.\cite{dftb+} Similarly, calculations using a density-functional ground state reference, which then uses the DFTB approximated random-phase excitation poles, becomes possible.\cite{10.1063/1.4948647} Generalization to spin-polarization or extending the real-time electronic propagation to receive an external model are also interesting further applications. Another extension built on top of the current work would be to exchange derivatives of the external models with respect to atomic displacements, enabling forces/strains from the Hellmann-Feynman theorem, or higher-order response properties using the internal \codename{DFTB+} coupled perturbed routines.\cite{maag2023mechanism}

In summary, the presented outcomes demonstrate the potential for flexible and powerful usage of components of the \codename{DFTB+} package by harnessing modularity. There is ample space for further integration of data workflows. The modularity of the package integration presents insertion points that can be used for evaluating a range of data objects in a variety of software packages, using the best implementations of any given step when these may be in separate software.



\textcolor{black}{
\section*{Supplementary Material}
The accompanying supplementary information provides: a comparison of band structures calculated with \codename{FHI-aims} and \codename{DFTB+}; details of the \codename{DFTB+} external API; details of the \codename{ASI} API.}

\section*{Data and Code Availability}
The \codename{DFTB+} software package is available at \url{https://github.com/dftbplus/dftbplus}. The v24.2 release will contain all functionality outlined in this manuscript; the described changes for the ASI binding or to connect to the \codename{ACEhamiltonians} are undergoing review and are currently available at~\cite{ASIbindings} and \cite{ACEbindings} respectively. Full documentation is available at \url{https://dftbplus.org/}.
The \codename{ACEhamiltonians} v0.1.0 software package is available at \url{https://github.com/ACEsuit/ACEhamiltonians.jl}.
The \codename{ASI} v1.1 software package \textcolor{black}{and \codename{DFTB+} implementation are} available at \url{https://gitlab.com/pvst/asi}. The interface specification \textcolor{black}{is} available at \textcolor{black}{\url{https://pvst.gitlab.io/asi}}. 

\section*{CRediT author statement}

\textbf{Pavel Stishenko}: Methodology, Software, Writing - Original Draft, Writing - Review \& Editing, Visualization.
\textbf{Adam McSloy}: Software, Visualization, Writing - Review \& 
Editing. 
\textbf{Berk Onat}: Software. 
\textbf{Ben Hourahine}: Conceptualization, Methodology, Writing - Original Draft, Writing - Review \& Editing, Visualization, Supervision, Project administration, Funding acquisition.  
\textbf{Reinhard J. Maurer}: Conceptualization, Writing - Original Draft, Writing - Review \& Editing, Supervision, Project administration, Funding acquisition.
\textbf{James Kermode}: Conceptualization, Methodology, Writing - Original Draft, Writing - Review \& Editing, Visualization, Supervision, Project administration, Funding acquisition.  
\textbf{Andrew Logsdail}: Conceptualization, Writing - Original Draft, Writing - Review \& Editing, Supervision, Project administration, Funding acquisition.

\begin{acknowledgments}

This work was financially supported by the Leverhulme Trust Research Project Grant (RPG-2017-191) and the NOMAD Centre of Excellence (European Commission grant agreement ID 951786). AJL and PVS acknowledge funding by the UKRI Future Leaders Fellowship program (MR/T018372/1, MR/Y034279/1). RJM acknowledges funding by the UKRI Future Leaders Fellowship program (MR/S016023/1, MR/X023109/1) and a UKRI frontier research grant (EP/X014088/1). PVS, RJM and AJL acknowledge funding from the ARCHER2 eCSE Programme (eCSE03-10).

We acknowledge computational resources provided by the Scientific Computing Research Technology Platform of the University of Warwick, Supercomputing Wales for access to the Hawk HPC facility, which is part-funded by the European Regional Development Fund via the Welsh Government, the EPSRC-funded HPC Midlands+ consortium (EP/T022108/1), and the UK Car-Parinello consortium (EP/P022065/1) and the U.K. High Performance Computing “Materials Chemistry Consortium” (EP/R029431/1, EP/X035859/1) for access to the ARCHER2 high-performance computing facility.

\end{acknowledgments}

\bibliography{JCP,paperpile}

\end{document}